\documentclass[onecolumn,groupedaddress,nofootinbib]{revtex4}
\usepackage{graphicx,amsfonts,amsmath,xcolor,epsfig,epstopdf}
\newcommand{\SU}[1]{\ensuremath{\mathrm{SU}( #1 )}}
\newcommand{\Un}[1]{\ensuremath{\mathrm{U}( #1 )}}

\newcommand{\SpR}[1]{\ensuremath{\mathrm{Sp}( #1,\mathbb{R} )}}
\newcommand{\betb}{\begin{tabular}{p{4.0cm}p{9.0cm}}}
\newcommand{\entb}{\end{tabular}}

\newcommand{\hw}{\ensuremath{\hbar\Omega}}
\newcommand{\ph}[1]{\ensuremath{#1}p-\ensuremath{#1}h}

\newsavebox{\smlmat}
\savebox{\smlmat}{$\left(\begin{smallmatrix} {\bf A} & \hspace{6pt} &  {\bf B} \\  {\bf C} & \hspace{6pt}  & {\bf D} \end{smallmatrix}\right)$}

\begin{document}

\title{Physics of nuclei: \\Key role of an emergent symmetry}
\author{T. Dytrych$^{1,2}$, K. D. Launey$^1$, J. P. Draayer$^1$,\\ D. Rowe$^3$, J. Wood$^4$, G. Rosensteel$^5$, C. Bahri$^4$, D. Langr$^6$, R. B. Baker$^1$}
\affiliation{$^1$Department of Physics and Astronomy, Louisiana State University, Baton Rouge, LA
70803}
\affiliation{$^2$ Nuclear Physics Institute, Academy of Sciences of the Czech Republic, 250 68 \u{R}e\u{z}, Czech Republic}
\affiliation{$^3$ Department of Physics, University of Toronto, Toronto, ON M5S 1A7, Canada}
\affiliation{$^4$ School of Physics, Georgia Institute of Technology, Atlanta, GA 30332}
\affiliation{$^5$ Physics Department, Tulane University, New Orleans, LA 70118}
\affiliation{$^6$Faculty of Information Technology, Czech Technical University in Prague, Praha, Czech Republic }

\maketitle

Exact symmetry and symmetry-breaking phenomena play a key role in providing a better understanding of the physics of many-particle systems, from quarks and atomic nuclei,  to molecules and galaxies.
In atomic nuclei, exact and dominant symmetries  such as rotational invariance, parity, and charge independence have been clearly established.
However, even when these symmetries are taken into account, the structure of nuclei remains illusive  and only partially understood, with no additional symmetries immediately evident from the underlying nucleon-nucleon  interaction.
Here, we show through {\it ab initio} large-scale nuclear structure calculations that the special nature of the strong nuclear  force determines 
additional highly regular patterns in nuclei that can be tied to an emergent approximate symmetry.
We find that  this symmetry is remarkably ubiquitous, regardless of its particular strong interaction heritage, and mathematically tracks with a symplectic group. 
Specifically, we show for light to intermediate-mass nuclei that the structure of a nucleus, together with its low-energy excitations,
 respects symplectic symmetry at about 70-80\% level, unveiling the predominance of only a few equilibrium shapes, deformed or not, with associated vibrations and rotations.
This establishes the symplectic symmetry as a remarkably good symmetry of the strong nuclear force, in the low-energy regime.
This may have important implications to studies, e.g., in astrophysics and neutrino physics that rely on nuclear structure information, especially where experimental measurements are incomplete or not available.
A very important practical advantage is that this new symmetry 
can be utilized to dramatically reduce computational resources required  in {\it ab initio} large-scale nuclear structure modeling. This, in turn, can be used  to pioneer predictions, e.g., for short-lived isotopes along various nucleosynthesis pathways. 

\section{Introduction}

The nucleus was discovered early in the last century.
Yet the nature of its dynamics  remains  poorly understood.
The standard shell model of nuclear physics is based on the premise that atomic nuclei have an underlying spherical harmonic oscillator (HO) shell structure,  as in the Mayer-Jensen  model \cite{MayerJ55},  with  residual interactions.
In fact, with effective interactions and large effective charges, the shell model is successful at explaining many properties of nuclei. 
However, it has been much less successful at predicting the many surprises that surface, such as the highly collective rotational states  that are observed and are described phenomenologically by the successful  Bohr-Mottelson collective model \cite{BohrM53},
as well as 
the recognition that the first excited state of the doubly closed shell  nucleus of $^{16}$O is the head  of a strongly deformed rotational band \cite{Morinaga56,RoweTW06}.
The coexistence of  states of widely differing deformation in many nuclei is now well established  \cite{HeydeW11,RoweW2010book,Wood16,Rowe16,RoweW18}
as an emergent phenomenon and dramatically exposes the limitations of the standard shell model.

To address this and to understand the physics of nuclei without limitations within the interaction and approximations during the many-body nuclear simulations, we use an {\it ab initio} framework that starts
with realistic interactions
tied to elementary particle physics considerations and fitted to nucleon-nucleon data.
Such calculations are now possible and are able to give 
converged results  for light nuclei by the use of  supercomputers.
However, in {\it ab initio} calculations the complexity of the nuclear problem dramatically increases with the number of particles, and when expressed in terms of literally billions of 
shell-model basis states, the structure of a nuclear state is unrecognizable.
But expressing it in a more informative basis, the
symmetry-adapted collective basis \cite{DytrychSBDV_PRL07,LauneyDD16}, leads to a major breakthrough: in
 this article, we report  on the very unexpected outcome from first-principle investigations of light to intermediate-mass nuclei (below the calcium region), namely, the incredible simplicity
  of nuclear low-lying states  
and the dominance we observe of an associated  symmetry of nuclear dynamics, the symplectic \SpR{3} symmetry, which  together with its slight symmetry breaking is shown here to naturally describe atomic nuclei.
This exposes for the first time the fundamental role of the symplectic \SpR{3} symmetry  and unveils it as a remarkably good symmetry of the strong nuclear force, represented here by interactions derived in the state-of-the-art chiral effective field theory.

\begin{figure}[th]
\centerline{
\includegraphics[height=0.30\textwidth]{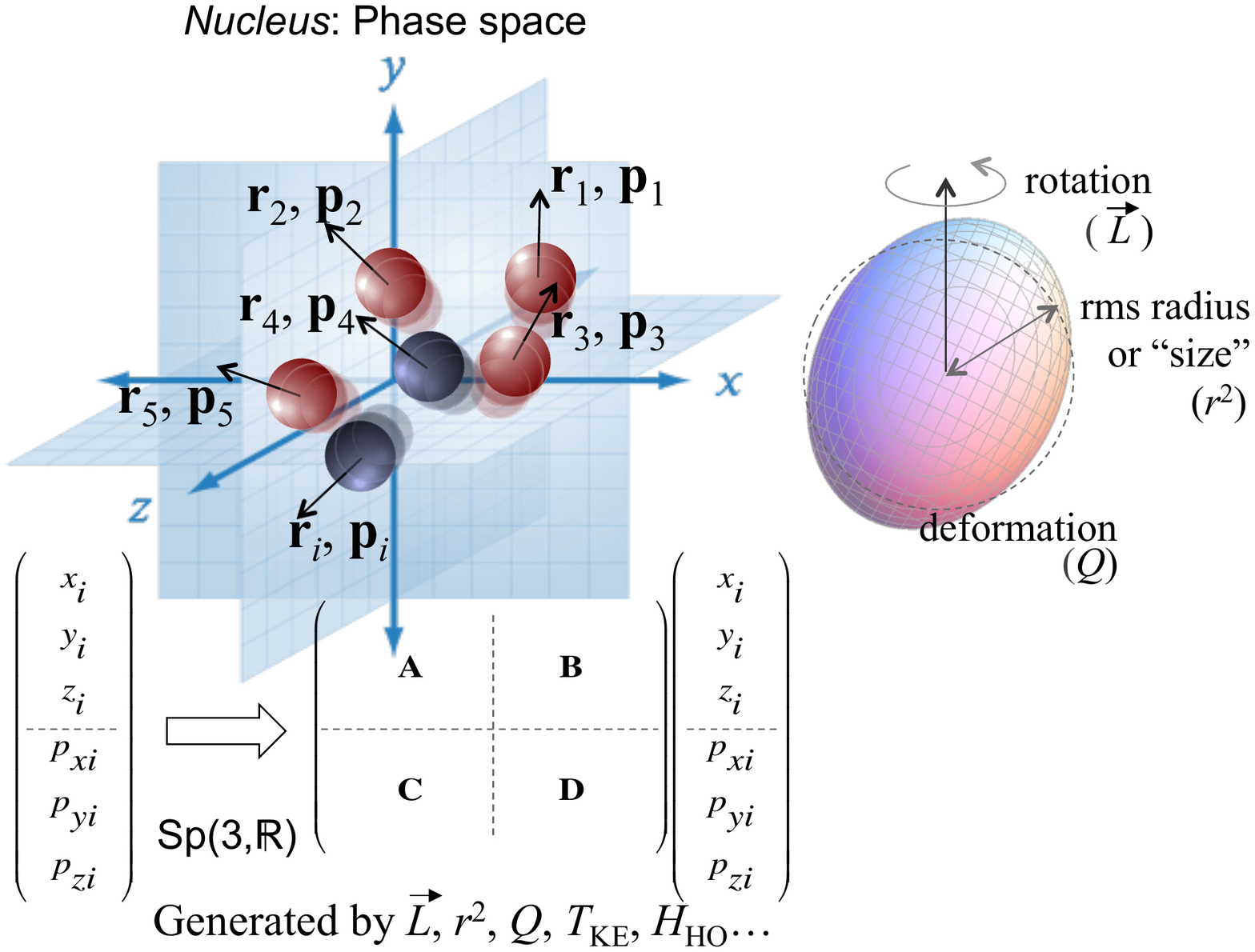}\hspace{32pt}
\includegraphics[height=0.30\textwidth]{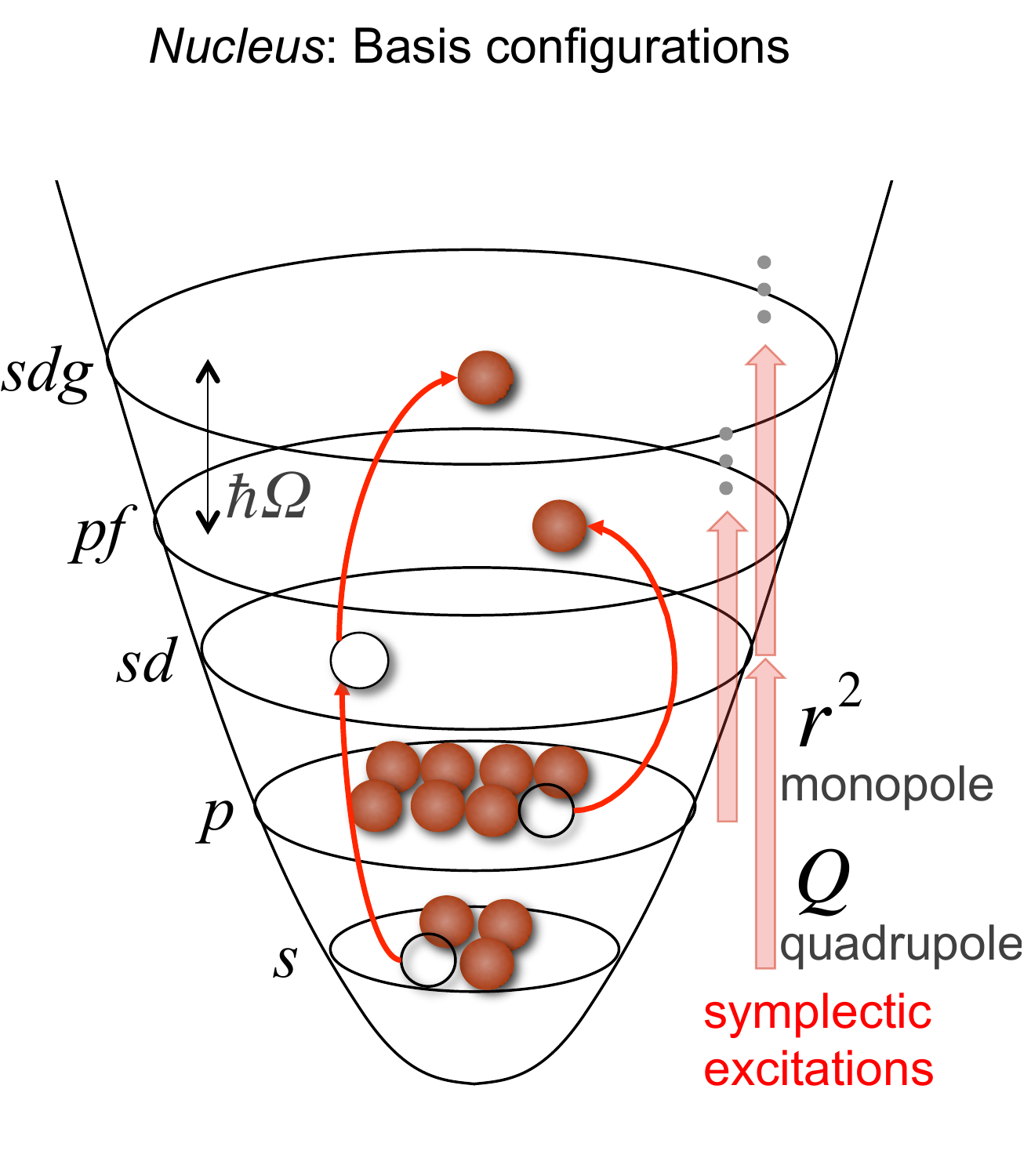}
}
\centerline{(a) \hspace{2.8in} (b)}
\caption{ 
(a) For $A$ particles in three-dimensional (3D) space, the complete basis for the translationally invariant shell model is described by \SpR{3(A-1)}$\times$\Un{4}, where \SpR{3(A-1)} is the group of all linear canonical transformations of the $3A$-particle phase space (excluding the spurious center-of-mass motion) and \Un{4} describes spin-isospin coordinates. The symplectic group \SpR{3} consists of all {\it particle-independent} linear canonical transformations 
\usebox{\smlmat}
of the single-particle phase-space observables $\vec r_i$ (position) and $\vec p_i$ (momentum) that preserve the Heisenberg commutation relations $[r_{\alpha i},p_{\alpha' j}]={\mathrm i}\hbar \delta_{\alpha \alpha'}\delta_{ij}$, $\alpha,\alpha'=x,y,z$ \cite{Rowe16}; generators of these transformations are expressed in ``quadratic coordinates" in phase space, $\sum_{i=1}^A r_{\alpha i} r_{\alpha^\prime  i}$, $\sum_{i=1}^A r_{\alpha i} p_{\alpha' i}\pm p_{\alpha i} r_{\alpha' i}$, $\sum_{i=1}^A p_{\alpha i} p_{\alpha' i}$, and include physically relevant operators:  orbital momentum $\vec L=\sum_{i}\vec r_i \times \vec p_i$, monopole moment $r^2=\sum_{i}{\vec r_i\cdot \vec r_i}$, quadrupole moment $Q_{2m}=\sqrt{16\pi/5 }\sum_i r_i^2Y_{2m}(\hat r_i)$,  
total kinetic energy $T_{\rm KE}=\frac{1}{2}\sum_{i}{\vec p_i\cdot \vec p_i}$, and the many-body HO  Hamiltonian $H_{\rm HO}=T_{\rm KE}+\frac{r^2}{2}$. 
(b) In the shell model, basis configurations are enumerated by monopole and quadrupole symplectic excitations. A key feature is that a single-particle \SpR{3} irreducible representation spans all positive-parity (or negative-parity) states for a particle in a 3D spherical or triaxial (deformed) harmonic oscillator.
}
\label{sp3Rphasespace}      
\end{figure}

It is known that SU(3), a subgroup  of \SpR{3}, is the symmetry group of the spherical harmonic oscillator that underpins the 
shell model \cite{MayerJ55} and the valence-shell \SU{3} (Elliott) model  \cite{Elliott58}.  
The Elliott  model has been shown  to naturally describe rotations of a deformed nucleus without the need for breaking rotational symmetry. 
The key role of deformation in nuclei and the coexistence of low-lying quantum states in a single nucleus characterized by configurations with different quadrupole moments 
\cite{HeydeW11,RoweW18} makes the quadrupole moment a dominant fundamental property of the nucleus, and together with the monopole moment or ``size" of the nucleus -- along with nuclear masses -- establishes the energy scale of  the nuclear problem. Indeed, the nuclear monopole  and quadrupole moments underpin the essence of  symplectic \SpR{3} symmetry (Fig. \ref{sp3Rphasespace}).  
Not surprisingly, the symplectic \SpR{3} symmetry, 
the underlying symmetry of the symplectic rotor model \cite{RosensteelR77,Rowe85}, 
has been found to play a key role across the nuclear chart -- from the lightest systems  \cite{RoweTW06,DreyfussLTDB13}, through intermediate-mass nuclei \cite{DraayerWR84,TobinFLDDB14,LauneyDD16}, up  to strongly deformed nuclei of the rare-earth and actinide regions \cite{Rowe85,CastanosHDR91,JarrioWR91,BahriR00}.
The results agree with experimental evidence that supports formation of enhanced deformation  and clusters in nuclei, as well as  vibrational and rotational patterns, as suggested by energy spectra, electric monopole and  quadrupole transitions, radii and quadrupole moments. And while these earlier symmetry-guided models have been very successful in explaining the observed collective patterns, they
have assumed symmetry-based approximations.  
Only now, an {\it ab initio} study -- without {\it a priori} assumptions of the symmetry and within a complete framework, using the symmetry-adapted no-core shell model (SA-NCSM) \cite{DytrychSBDV_PRL07,DytrychLMCDVL_PRL12,LauneyDD16} --  shows that dominant features of nuclei track with the symplectic symmetry and naturally
emerge from first-principle  considerations,
even in close-to-spherical nuclear states without any recognizable rotational properties. Therefore, the present outcome not only explains but also predicts the emergence of nuclear collectivity.

\section{Method}

The {\it ab initio} nuclear shell-model theory \cite{CaurierMNPZ05,BarrettNV13} solves the many-body Schr\"odinger equation for $A$ particles,
\begin{equation}
H \Psi(\vec r_1, \vec r_2, \ldots, \vec r_A) = E \Psi(\vec r_1, \vec r_2, \ldots, \vec r_A), {\rm with}\, H = T_{\rm rel} + V_{NN}  + V_{3N} + \ldots + V_{\rm Coulomb},
\label{ShrEqn}
\end{equation}
and, in its most general form, is an exact many-body ``configuration interaction" method, for which the interaction and basis configurations are as follows. 
\newline
{\bf Interaction --} 
The intrinsic non-relativistic nuclear  Hamiltonian $H$ includes
the relative kinetic energy $T_{\rm rel} =\frac{1}{A}\sum_{i<j}^A\frac{(\vec p_i - \vec p_j)^2}{2m}$ ($m$ is the nucleon mass),  
the nucleon-nucleon ($NN$) and, possibly, three-nucleon ($3N$) interactions, 
along with the Coulomb interaction between the protons.  In our study, we have adopted various realistic interactions without renormalization in nuclear medium (referred to as ``bare"), with results illustrated here for up to the next-to-next-to-next-to-leading order (fourth order), namely, for the Entem-Machleidt (EM) N$^3$LO \cite{EntemM03} and  NNLO$_{\rm opt}$ \cite{Ekstrom13} chiral potentials.  
We neglect explicit $3N$ interactions, since they are known to be hierarchically smaller than $NN$ and to contribute only slightly to densities. In addition,  the symmetry patterns for the EM-N$^3$LO  $NN$ interaction exhibit a surprisingly similar  behavior, as shown below, as the ones for the   NNLO$_{\rm opt}$  $NN$ interaction, which minimizes $3N$ contributions in $^3$H and $^{3,4}$He \cite{Ekstrom13} as compared to other parameterizations of chiral  interactions up to N$^3$LO.
\newline
{\bf Basis configurations and symmetry-adapted (SA) basis --} 
A complete orthonormal basis $\psi_k$ is adopted, such that the expansion 
$\Psi(\vec r_1, \vec r_2, \ldots, \vec r_A) = \sum_{k} c_k \psi_k(\vec r_1, \vec r_2, \ldots, \vec r_A)$,
 renders Eq. (\ref{ShrEqn}) into a matrix eigenvalue equation with unknowns $c_k$,
$
\sum_{k'} H_{k k'} c_{k'} = E c_k,
$
where the many-particle Hamiltonian matrix elements
$H_{k k'} = \langle \psi_k | H | \psi_{k'} \rangle$ are calculated for the given interaction (the solution $\{c_k^2\}$ defines a set of probability amplitudes). Here, the basis is a finite set of antisymmetrized products of  single-particle states of a spherical harmonic oscillator 
of frequency $\hbar\Omega$, referred to as a ``model space", that is truncated by the total number of HO quanta $N_{\rm max}$. With larger model spaces (higher $N_{\rm max}$), the  eigenenergies and nuclear observables become independent of \hw~ and converge to the exact  values. Such a basis allows for preservation of translational invariance of the nuclear self-bound system. 
Furthermore, the model space can be reorganized via a unitary transformation -- without loss of information -- to a basis that respects an approximate symmetry of the nuclear system, referred to as a symmetry-adapted basis. 
This leads to a  much faster convergence and to quantum states that can be described by a drastically smaller number of SA basis states.

\vspace{-11pt}
\section{Results and discussions}
\vspace{-9pt}
\subsection{Nature's preference}
\vspace{-10pt}
In this article, we report on the remarkable outcome, as unveiled from first-principle calculations of nuclei below the calcium region,
 that nuclei exhibit relatively simple physics. We now understand that a low-lying nuclear state
is predominantly composed 
of a few equilibrium shapes that vibrate and rotate, with each shape 
characterized
by a single symplectic irreducible representation (irrep). 
\begin{figure}[th]
\begin{minipage}{0.68\textwidth}
\includegraphics[width=\textwidth]{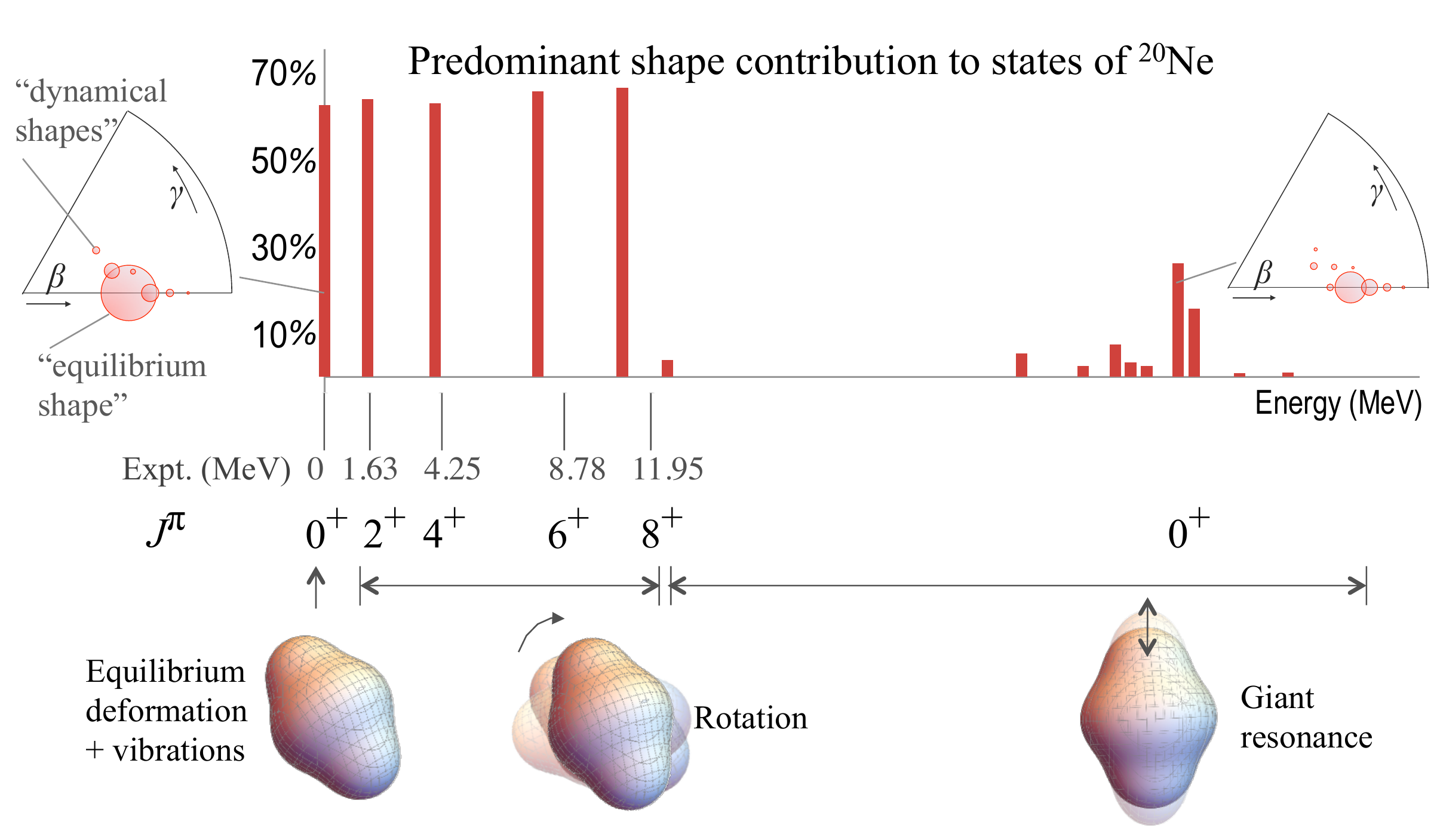}\\
\end{minipage}
\begin{minipage}{0.31\textwidth}
\includegraphics[width=0.5\textwidth]{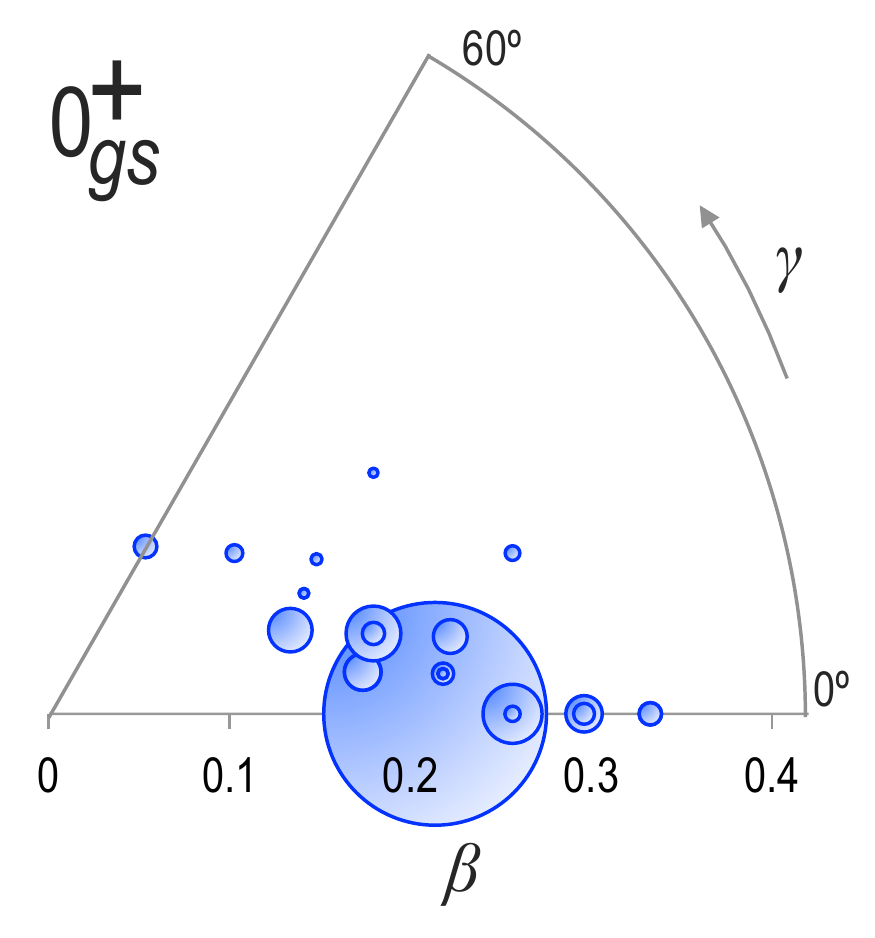} \\
\includegraphics[width=0.5\textwidth]{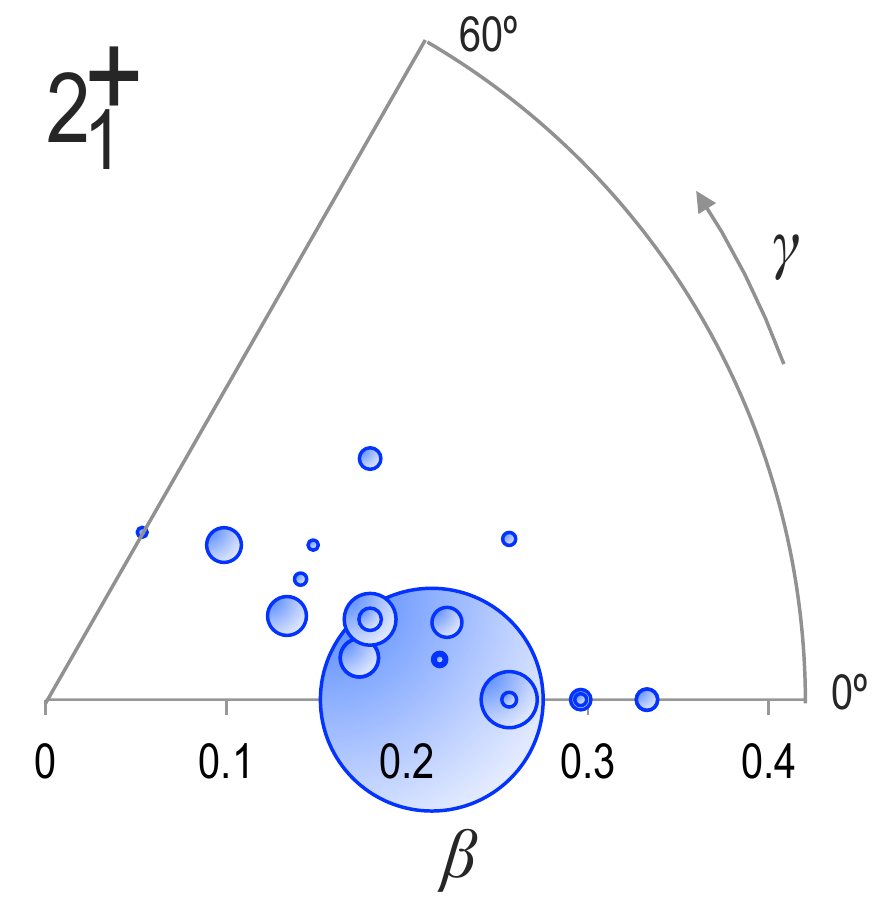} \\
\end{minipage}   \\
\hspace{1.5in}(a) \hspace{3.1in} (b)
\caption{ \label{en_shapes_vibr_rot}   
(a) Excitation energies (horizontal axis) of the ground-state ({\it gs}) rotational band ($J^\pi=0^+,2^+,4^+,6^+$, and $8^+$)  and $0^+$ states in $^{20}$Ne,  shown together with the contribution to each state (vertical axis) of the  single shape 
that dominates the ground state. According to this, states are grouped  and schematically illustrated by ``classical" shapes, vibrations, and rotations, where the {\it ab initio} one-body density profile in the {\it body-fixed} frame  is shown for $0^+$. (b) Deformation distribution of  the equilibrium shapes that make up a state  with contributions given by the area of the circles, 
specified by 
the average deformation $\beta$ and triaxiality angle $\gamma$.
Results reported for {\it ab initio} SA-NCSM calculations  with the bare NNLO$_{\rm opt}$ $NN$ for an \SU{3}-adapted basis that yields a fast convergence of the  {\it gs} rms radius (model space of 11 HO shells with 15 MeV inter-shell distance). 
}
\end{figure}

To illustrate this, we consider the physics  of $^{20}$Ne (Fig. \ref{en_shapes_vibr_rot}) and the contribution of a single symplectic irrep to its low-lying states, Fig. \ref{en_shapes_vibr_rot} (a).
Indeed, the physics of a single symplectic irrep can provide insight into the nuclear dynamics: all configurations within a symplectic irrep preserve an equilibrium shape and realize its rotations, vibrations, and spatial orientations,  implying that the $^{20}$Ne {\it ab initio}  wave functions for $J^\pi=0^+_{gs},2^+,\dots, 8^+$ indeed exhibit a predominance of a single  equilibrium shape that vibrates and rotates [see also, Fig. \ref{en_shapes_vibr_rot} (b), largest circle]. That a single  \SpR{3} irrep naturally describes the shape dynamics of a deformed nucleus
has been earlier shown in an algebraic symplectic model \cite{RosensteelR77,Rowe85,BahriR00} and can be illustrated with two simple examples: (1) in the limit of a valence shell, 
the  symplectic basis recovers the \SU{3}-adapted basis of the Elliott model that describes shapes (referred to as ``equilibrium shapes" or simply ``shapes") and their rotations  -- note that a shape associated with a given many-body \SU{3}-adapted state  is specified by the familiar shape parameters,  $\beta$ and  $\gamma$, calculated according to the  correspondence of the expectation value of $Q\cdot Q$ and $Q\times Q \cdot Q$  to $\beta^2$ and $\beta^3\cos{3\gamma}$, respectively \cite{CastanosDL88}; and (2) for a single spherical shape, its symplectic excitations
(referred to as ``dynamical shapes") 
realize the microscopic counterpart of the surface vibrations of the Bohr-Mottelson collective model \cite{Rowe13}.
As further shown in the $\beta$-$\gamma$ plots of Fig. \ref{en_shapes_vibr_rot}(a), the set of higher-lying $0^+$ states with nonnegligible contribution of the \ph{1} vibrations of the ground-state shape describes a fragmented giant monopole resonance (breathing mode) with a centroid around $29$ MeV  and a typical deformation content spread out to higher $\beta$ values due to vibrations \cite{BahriDCR90}, as compared to the ground state. 
Implications of this outcome for understanding the physics of nuclei are discussed next.
\begin{figure}[th]
\begin{tabular}{ll}
{\footnotesize (a) $^6$Li, EM-N$^3$LO $NN$} & {\footnotesize(b) $^8$He, NNLO$_{\rm opt}$ $NN$}\\
\includegraphics[width=0.47\textwidth]{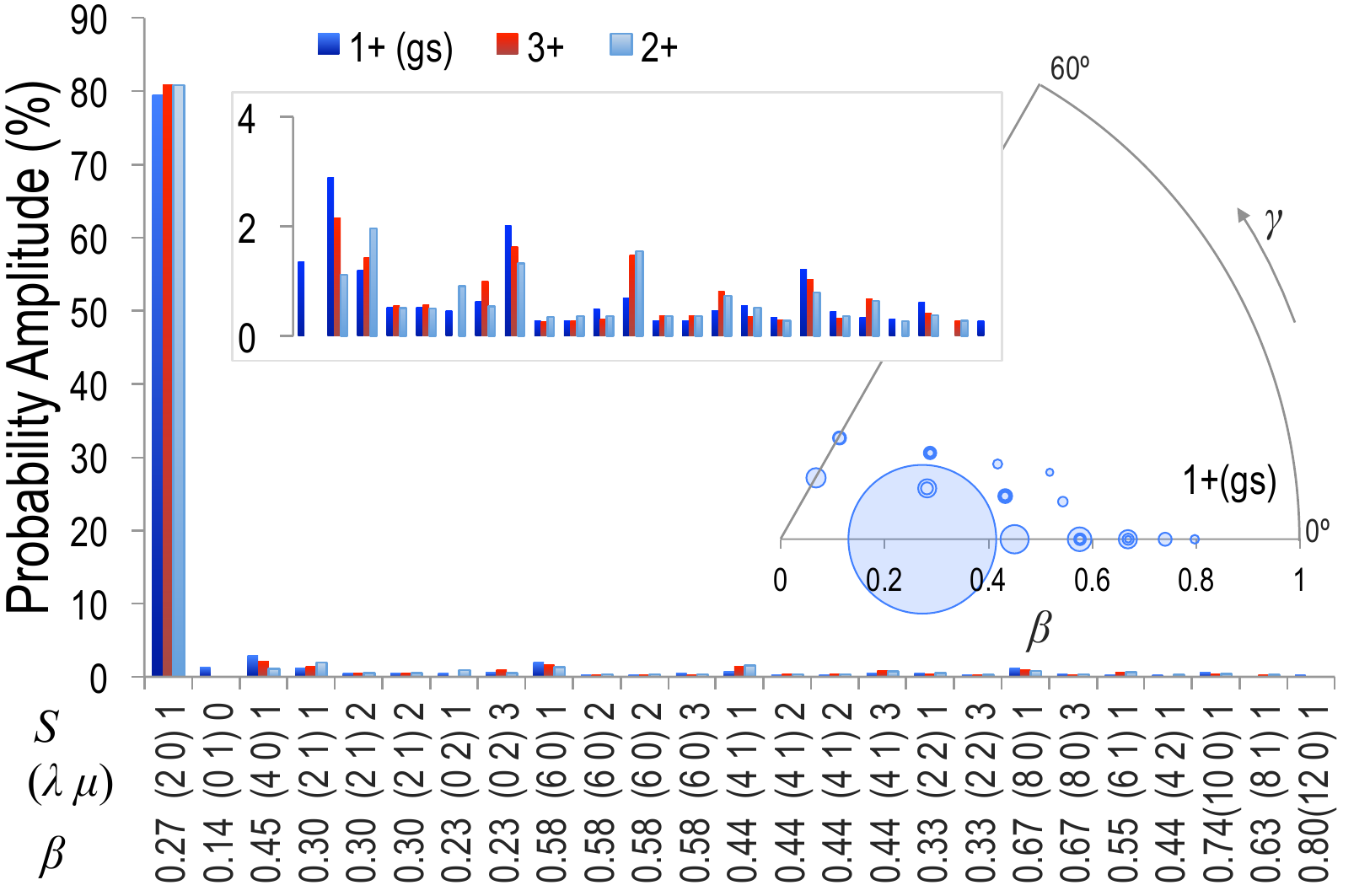} &
\includegraphics[width=0.45\textwidth]{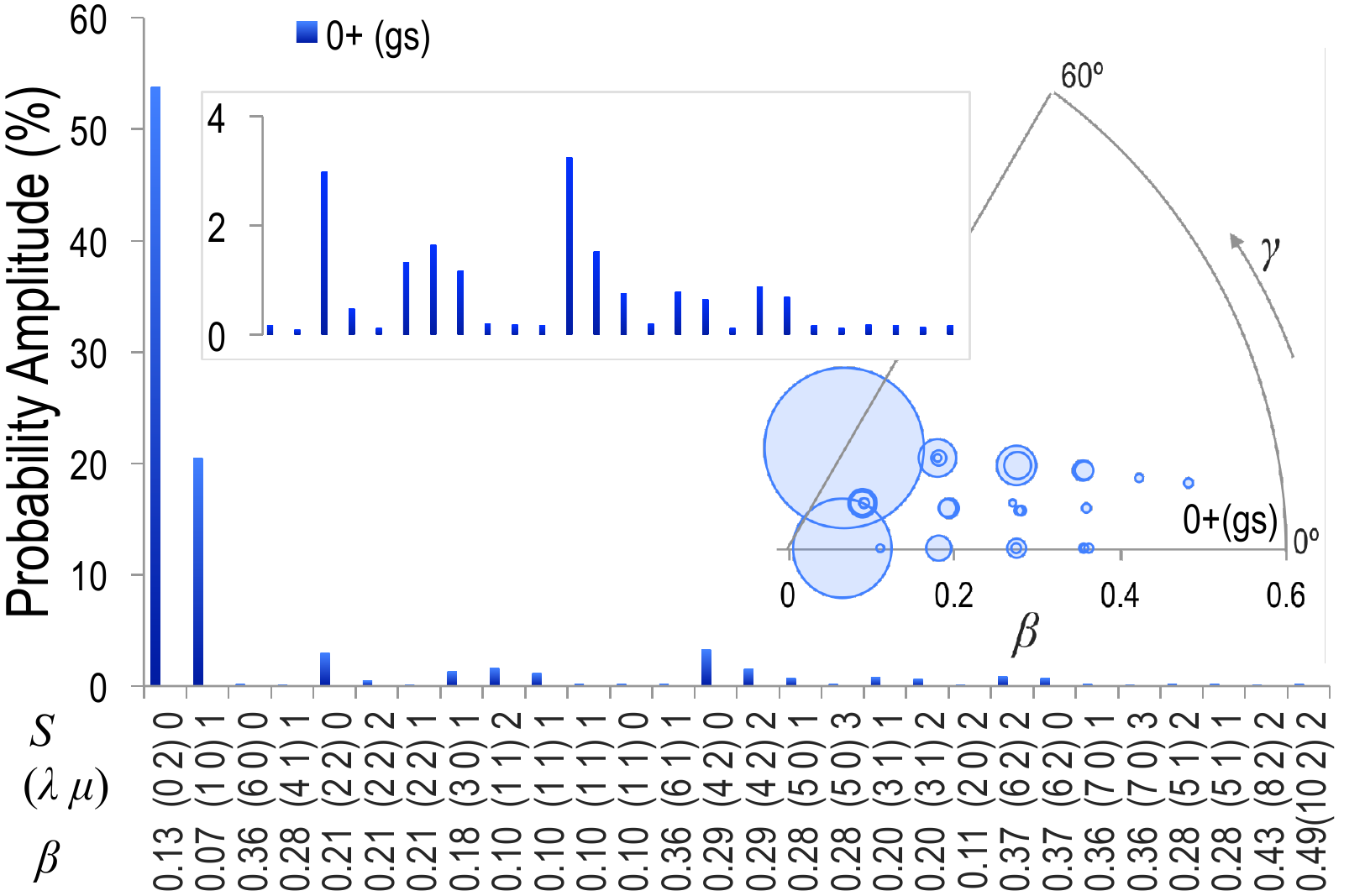} \\
(c) {\footnotesize$^{20}$Ne, EM-N$^3$LO $NN$} & {\footnotesize (d) NNLO$_{\rm opt}$ $NN$}\\\
\includegraphics[width=0.47\textwidth]{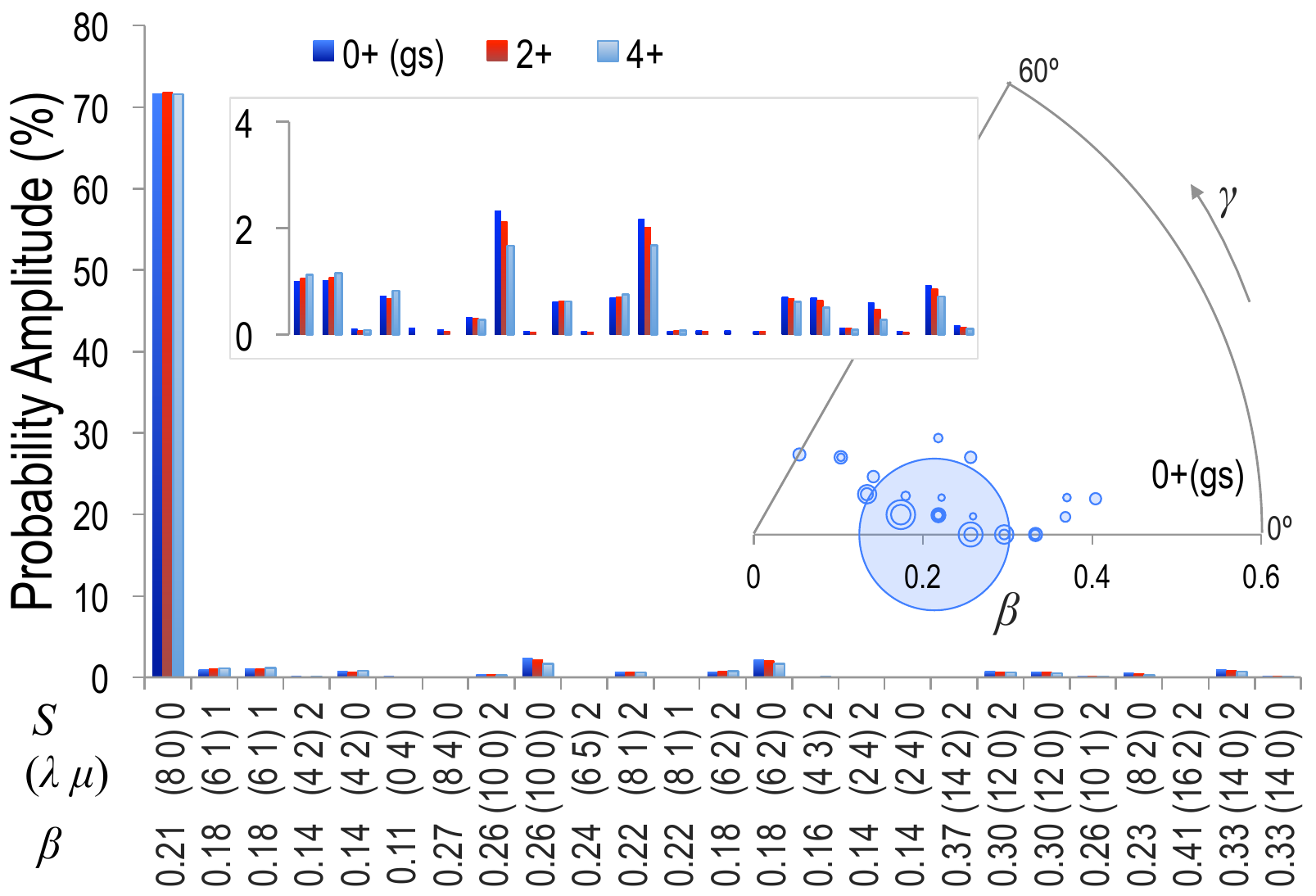}&
\includegraphics[width=0.24\textwidth]{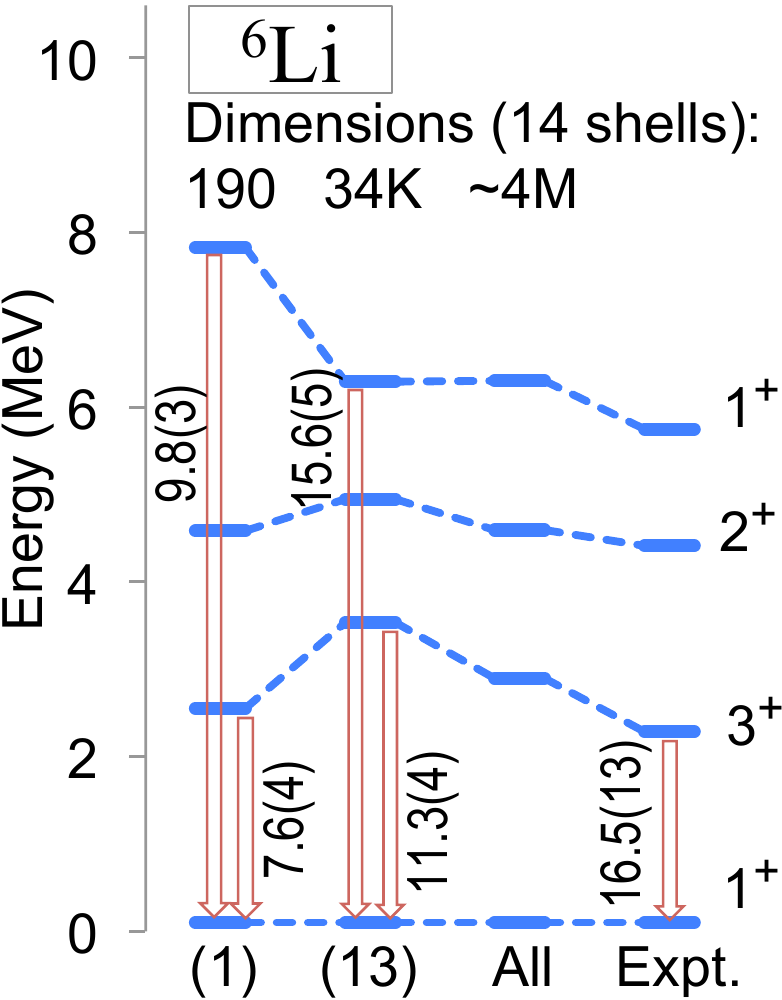}
\includegraphics[width=0.24\textwidth]{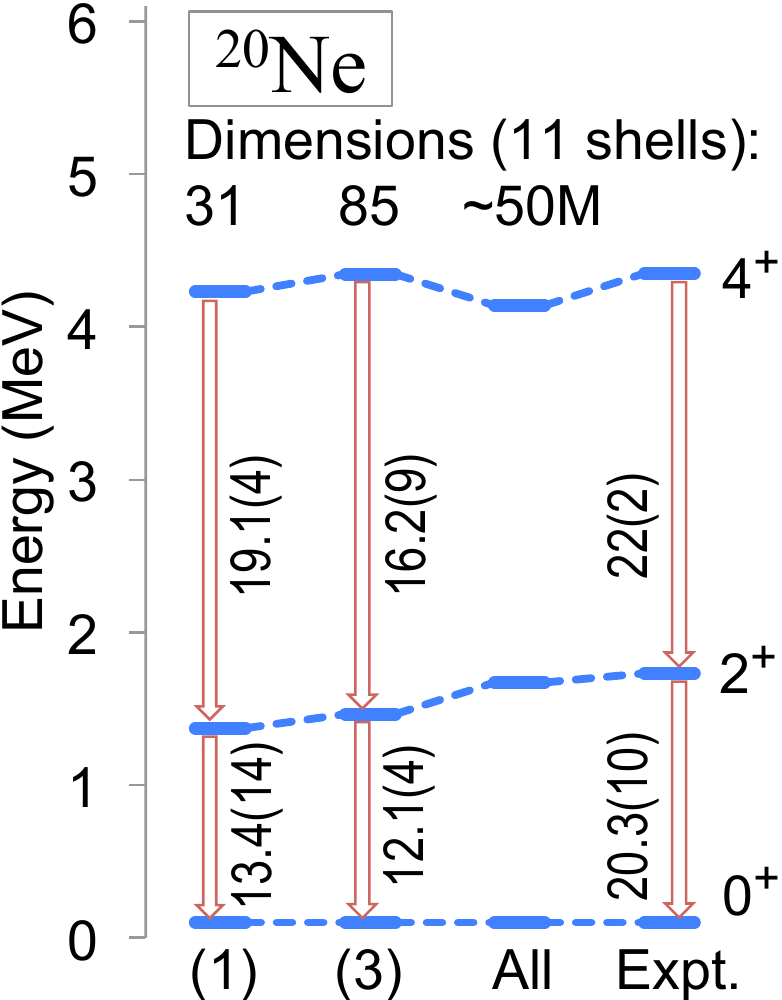}
\end{tabular}
\caption{(a)-(c) Symplectic  \SpR{3} irreps  that make up the rotational band  states of $^6$Li, $^{8}$He, and $^{20}$Ne (in a close agreement with  the results of Fig. \ref{en_shapes_vibr_rot});  each irrep is specified by its equilibrium shape, labeled by the shape deformation $\beta$ and the  corresponding \SU{3} labels $(\lambda\,\mu)$ together with total intrinsic spin $S$.
Insets:  the same irreps but without the predominant  contribution, together with the $\beta$-$\gamma$ plot for the ground state.  
(d) Observables for $^{6}$Li  and $^{20}$Ne calculated   in the  {\it ab initio} SA-NCSM with \SpR{3}-adapted basis  using only a small number (specified in the $x$-axis labels) dominant symplectic irreps including the most dominant  one,
as compared to experiment (``Expt."); dimensions of the largest model spaces used  are also shown. 
Energies (with errors $\sim 100$ keV) and reduced electromagnetic $B(E2)$ transition strengths (in W.u.) are reported  for extrapolations to infinitely many  shells of converging results  across variations in the model space size and resolution (see also Fig. \ref{6Li_N3LOcoul_sp_IR_UB}).
Results for nuclear states (a)-(c) and energies [(d), labeled as ``All"] are reported for {\it ab initio} SA-NCSM calculations for an \SU{3}-adapted basis  that yields a fast convergence of the  {\it gs} rms radius: complete (selected) model space of 14 (11)  HO major shells for $^6$Li and $^{8}$He ($^{20}$Ne) with inter-shell distance of (a)-(b) 20 MeV and (c)-(d) 15 MeV.
}
\label{sp_Li6_Ne20}      
\end{figure}

\vspace{-10pt}
\subsection{Approximate symplectic symmetry and physics of nuclei }
\vspace{-10pt}
The \SpR{3}-adapted basis is constructed  for various nuclei, pointing to unexpectedly ubiquitous symplectic symmetry, with the illustrative examples for the odd-odd $^6$Li, $^8$He (generally considered to be spherical),  and the intermediate-mass $^{20}$Ne shown in Fig. \ref{sp_Li6_Ne20}. 
The outcome provides further evidence that nuclei are predominantly comprised -- typically in excess of 70-80\% -- of only  a few 
 shapes, often a single shape (a single symplectic irrep) as for, e.g.,  $^{6}$Li, $^{8}$B, $^{8}$Be, $^{16}$O, and $^{20}$Ne, or two shapes, e.g., for $^{8}$He and $^{12}$C [see also results in Ref. \cite{LauneyDD16} based on \SU{3} analysis]. Hence, e.g., the ground sate of $^{6}$Li and $^{20}$Ne ($^{16}$O) is found to exhibit prolate (spherical) shape deformation,  while an oblate shape dominates in the case of $^{8}$He. 
Besides the predominant irrep(s), there is a manageable number of symplectic irreps, each of which contributes at a level that is typically at least an order of magnitude  smaller, as shown in  Fig.  \ref{sp_Li6_Ne20}(a)-(c).
Furthermore, the outcome implies that the richness of the low-lying excitation spectra naturally emerges from these shapes through their rotations. Indeed, practically the same symplectic content observed for the low-lying states in $^6$Li, Fig. \ref{sp_Li6_Ne20}(a), and for those in $^{20}$Ne, Fig. \ref{sp_Li6_Ne20}(c), is a rigorous signature of rotations of a shape and can be used to identify members of a rotational band. 
And finally, 
$E2$ transitions are determined by the quadrupole operator $Q$, an  \SpR{3} generator that does not mix symplectic irreps -- the predominance of a single symplectic irrep reveals the remarkable result that the largest fraction of these transitions, 
and hence nuclear collectivity,  necessarily emerges within this symplectic irrep, Fig. \ref{sp_Li6_Ne20}(d) [similarly for rms radii, since $r^2$ is also an  \SpR{3} generator]. A notable outcome is that even excitation energies calculated in model spaces selected down to a few symplectic irreps closely reproduce the experimental data.

\begin{figure}[th]
\includegraphics[width=0.56\textwidth]{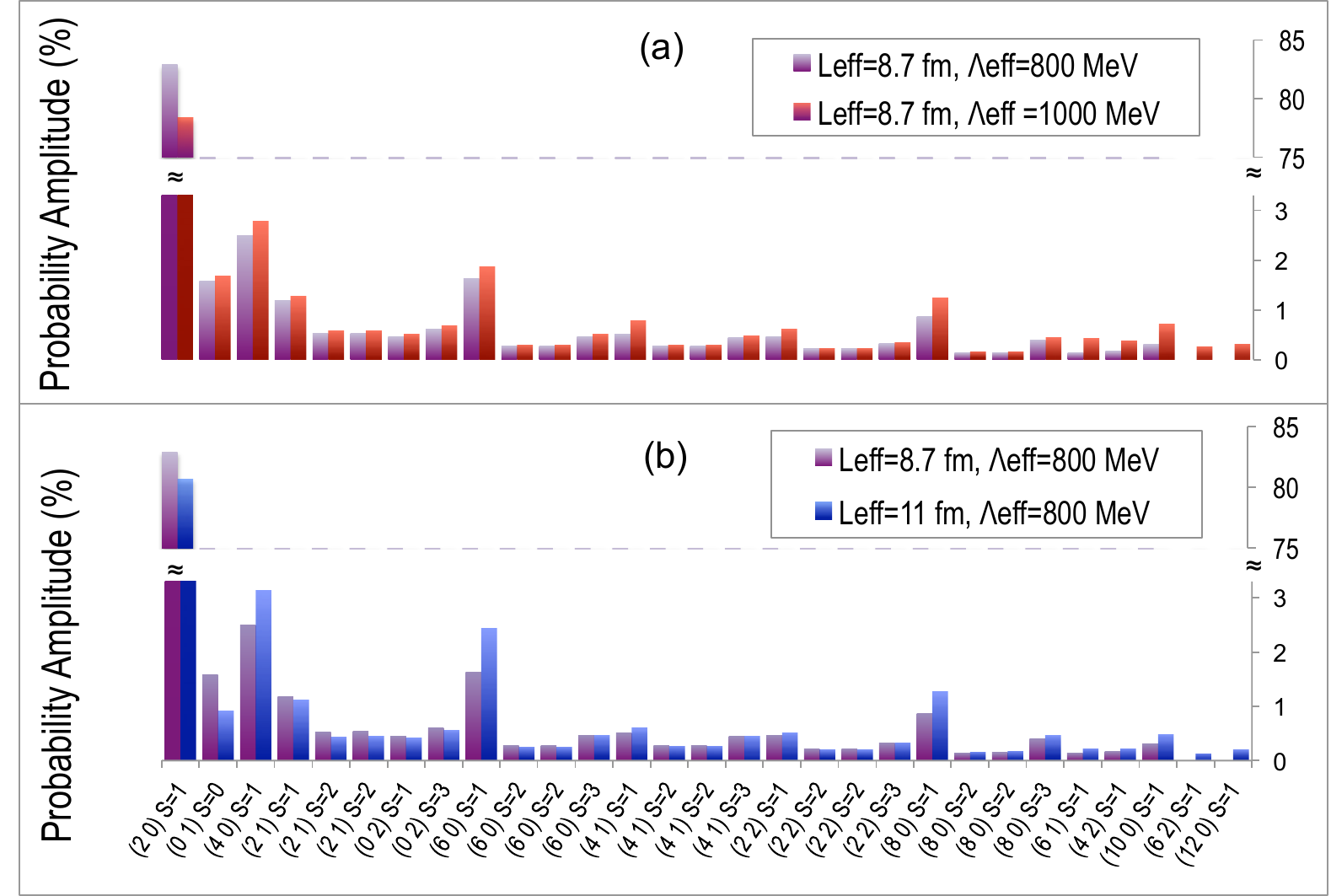}
\caption{Symplectic  \SpR{3} irreps, labeled by $(\lambda\,\mu)S$, that make up the ground sate of $^6$Li, as calculated by the {\it ab initio} SA-NCSM with \SU{3}-adapted basis with the bare N$^3$LO $NN$ interaction and the effect on the symplectic content (a) as  the resolution improves (grid size decreases) for the same box size, and  (b) as the box size increases for the same resolution. No new dominant equilibrium shapes are observed as the box size or grid resolution increases.
}
\label{6Li_N3LOcoul_sp_IR_UB}      
\end{figure}

The outcome is neither sensitive to the type of the realistic interaction used (details such as contribution percentages slightly vary, but dominant features retain), nor to the parameters of the basis, \hw~ and $N_{\rm max}$. It has been shown \cite{WendtFPS15} that these two model parameters can be related to $L_{\rm eff}$, the infrared IR cutoff,  and $a_{\rm eff}$, the ultraviolet UV cutoff $\Lambda_{\rm eff}=1/a_{\rm eff}$, which can be understood as the effective size of the model space box in which the nucleus resides and its grid size (resolution), respectively. Indeed, the symplectic content of a nucleus
is found to be stable against variations in  the box size or resolution -- Fig. \ref{6Li_N3LOcoul_sp_IR_UB} reveals that no new dominant 
shapes appear for values around the optimal ones for $^6$Li shown in Fig. \ref{sp_Li6_Ne20}(a), retaining the predominance of the single irrep. This has an important implication: complete SA-NCSM calculations are performed in smaller box sizes and/or low resolution to identify the nonnegligible symplectic irreps, while the model space is then augmented by extending these irreps to high (otherwise inaccessible) HO major shells vital to account for collectivity. 

In short, this work shows that nuclei below the calcium region and their low-energy excitations display relatively simple emergent physics that is collective in nature and tracks with an approximate symplectic symmetry heretofore gone unrecognized in the strong nuclear force.

\section{Acknowledgements}
This work was supported by the U.S. National Science Foundation (OIA-1738287, ACI -1713690) and the Czech Science Foundation(16-16772S), and SURA. This work benefitted from computing resources provided by
Blue Waters, LSU ({\tt www.hpc.lsu.edu}), and the National Energy Research Scientific
Computing Center (NERSC). The Blue Waters sustained-petascale computing project is supported by the National Science Foundation (awards OCI-0725070 and ACI-1238993) and the state of Illinois, and is a joint effort of the University of Illinois at Urbana-Champaign and its National Center for Supercomputing Applications. 

\bibliographystyle{unsrtnat}
\bibliography{lsu_latest}

\end{document}